\documentclass[conference]{IEEEtran}
\usepackage{graphicx}
\usepackage{comment}
\usepackage{subcaption} 
\usepackage[inline]{enumitem}
\usepackage{url}
\usepackage{flushend} 

\begin{document}

\title{OFDM Waveform for Monostatic ISAC in 6G:\\ Vision, Approach, and Research Directions}

\author{
    \IEEEauthorblockN{Huacheng Zeng, Kunzhe Song, Geo Jie Zhou, and Ruxin Lin}
    \IEEEauthorblockA{{Department of Computer Science and Engineering, Michigan State University}}
}
 
\maketitle

\begin{abstract}
Integrated sensing and communication (ISAC) is widely regarded as a key enabling technology for 6G wireless networks. While extensive research has explored the coexistence of sensing and communication functionalities, the use of orthogonal frequency-division multiplexing (OFDM) waveforms for monostatic ISAC remains underexplored.
In this article, we present practical approaches for enabling monostatic sensing on wireless communication devices and illustrate how OFDM signals can provide radar-like sensing capabilities such as ranging, Doppler estimation, and environmental perception. 
We hope this article will stimulate further research on OFDM-based monostatic ISAC and accelerate its adoption in 6G networks.
\end{abstract}


\section{Introduction} 

Integrated sensing and communication (ISAC) is emerging as a key capability of 6G wireless networks \cite{zhu2024enabling}. Over the past decade, a substantial body of research has explored ISAC techniques for wireless systems. In general, sensing and communication can be integrated at three levels: \emph{spectrum sharing}, \emph{hardware sharing}, and \emph{waveform sharing}. Early work primarily focused on spectrum sharing, where sensing and communication systems operate in the same frequency band but rely on separate hardware components. Recent research has moved toward deeper integration by jointly optimizing waveform, beamforming, and resource allocation for sensing and communication functions.
One example is channel state information (CSI)-based sensing in Wi-Fi networks \cite{ma2019wifi}. In this paradigm, the communication system itself serves as a sensing platform, with spectrum, hardware, and waveform shared between communication and sensing tasks.

Despite these advances, most existing ISAC systems operate under a \emph{bistatic} configuration, where the transmitter and receiver are located on different devices (see, e.g., \cite{naoumi2024complex,brunner2024bistatic}). In such systems, the receiver measures channel variations caused by environmental changes and infers sensing information from these variations. While bistatic sensing has enabled many applications such as activity recognition and gesture detection, it has several fundamental limitations. Because the transmitter and receiver are physically separated and typically unsynchronized in time and frequency, the measured channel response cannot directly provide absolute propagation delay or Doppler frequency. As a result, bistatic sensing often relies on heuristic algorithms or machine learning models to infer motion patterns rather than measuring physical quantities such as distance or velocity \cite{xu2026hybrid}. This limitation also reduces generalizability, as learned models may depend heavily on the specific scenes seen during training.

In contrast, \emph{monostatic} ISAC refers to the configuration in which a communication device performs sensing using its own transmitted signal, with the transmitter and receiver co-located on the same device \cite{keskin2025fundamental,bomfin2024performance}. This architecture enables coherent detection of reflected signals and allows direct estimation of physical parameters such as propagation delay and Doppler frequency. Consequently, monostatic ISAC can provide radar-like capabilities, including ranging, velocity estimation, and micro-Doppler sensing, while leveraging existing communication signals. Despite these advantages, monostatic ISAC remains underexplored in wireless communication research.

In this article, we advocate monostatic ISAC as an important direction for future 6G networks. Compared with bistatic approaches, monostatic ISAC allows individual communication devices to directly measure the channel impulse response (CIR) of their surrounding environments. When a device transmits an orthogonal frequency-division multiplexing (OFDM) communication signal, the reflected signals from surrounding objects are received by its own receiver. By estimating the channel response from the received signal, the device can recover the time-domain CIR, which directly reveals the propagation delays of different reflection paths. Because the transmitter and receiver share the same clock and frequency reference, these measurements can be used to estimate object distance and Doppler velocity, enabling radar-like sensing capabilities using standard communication waveforms.

A key challenge in realizing monostatic ISAC on communication devices is enabling the transmitter and receiver to operate simultaneously or in rapid alternation without overwhelming the receiver with self-interference from the transmitted signal. To address this challenge, we discuss four approaches that allow communication devices to perform monostatic sensing while maintaining normal communication functionality: 
mmWave Tx/Rx antenna isolation for mmWave devices, self-interference cancellation, self-mixing RF down-conversion, and precise Tx/Rx RF switching.
The first approach is suitable for mmWave systems, while the remaining three approaches are primarily designed for sub-6 GHz devices.

Building upon these techniques, we further discuss potential applications of monostatic ISAC in future wireless networks and highlight the role of artificial intelligence (AI) in interpreting RF sensing data. By enabling communication devices to directly perceive their environments, monostatic ISAC has the potential to transform 6G network infrastructure into a pervasive sensing platform. 

\section{Monostatic ISAC in 6G}

\begin{figure}[t]
\centering
  \includegraphics[width=\linewidth]{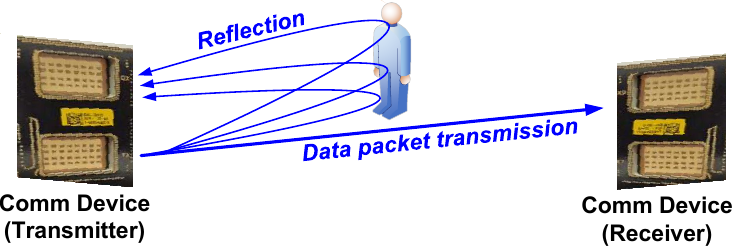}
  \caption{A communication device operates in full-duplex mode. It is transmitting OFDM signal for data packet transmission; at the same time, its receiver demodulates the OFDM symbols to estimate CIR, which is used for sensing applications.}
  \label{fig:cir1}
\end{figure}

\subsection{Single Device for Monostatic Sensing}

OFDM has been dominantly used in existing communication systems such as 4G/5G and Wi-Fi, and it will continue to be used in 6G systems.
As illustrated in Fig.~\ref{fig:cir1}, when a communication device transmits an OFDM signal, the emitted radio waves illuminate the surrounding environment. Objects in the environment reflect part of the signal back toward the device, and the receiver on the same device can capture these reflections. Using the received signal, the device can estimate the channel response of individual OFDM subcarriers and convert it to the CIR.
The CIR measurement, $\vec{h} = [h_0, h_1, \ldots, h_{L-1}]$, encodes sensing information about the environment. The delay resolution of the CIR taps is $c/(2B)$, where $c$ is the speed of light and $B$ is the OFDM signal bandwidth. A large value of $h_l$ indicates a high likelihood of an object located at a distance of $lc/(2B)$. 

If the transmitter and/or receiver is equipped with multiple antennas arranged along horizontal and vertical dimensions, the antenna array can be used to estimate the angular information of reflected signals. By combining ranging and angular measurements, a 3D radio point cloud of the surrounding environment can be generated for downstream sensing tasks. Furthermore, when multiple point cloud frames are collected over time, the phase variation of individual radio points can be used to estimate Doppler frequency shifts. This enables the construction of 4D sensing data, including range, azimuth, elevation, and Doppler information.

During the sensing process, self-interference leakage from the transmitter to the receiver appears as a static component in the measured CSI and CIR. Because this leakage corresponds to a fixed channel path, it can be modeled as a constant term and removed during signal processing.
However, although self-interference can be removed relatively easily in the digital domain, excessive self-interference power may saturate the ADC, causing the signal of interest to be severely affected by quantization and thermal noise.
To enable monostatic ISAC on individual devices, future 6G systems need to support full-duplex or quasi-full-duplex operation to effectively manage self-interference, allowing devices to transmit and receive signals simultaneously or in rapid alternation.



\subsection{Understanding RF Sensing Data}

For a device equipped with monostatic ISAC capabilities, a fundamental question is what types of sensing data can be extracted from communication signals and how these data are represented. Modern radar systems typically produce 4D sensing information, including range, Doppler velocity, azimuth angle, and elevation angle. In fact, monostatic ISAC systems can obtain comparable 4D sensing information from the channel measurements of OFDM signals.

\begin{itemize}[leftmargin=0.15in]

\item 
\textbf{Ranging:}  
In a monostatic setting, the transmitter and receiver are co-located on the same device and share synchronized time and frequency references. 
Therefore, the measured CIR directly reflects the propagation delay of signal reflections, which corresponds to object distance.
The ranging resolution of the system is
\(
\Delta R = \frac{c}{2B}
\).


\begin{figure}
    \centering
    \includegraphics[width=0.4\linewidth]{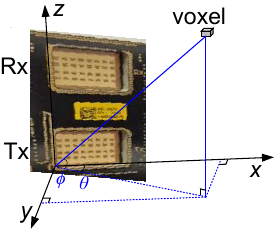}
    \caption{Angular estimation on a communication device.}
    \label{fig:cir2}
\end{figure}

\begin{figure*}
    \centering
    \includegraphics[width=\linewidth]{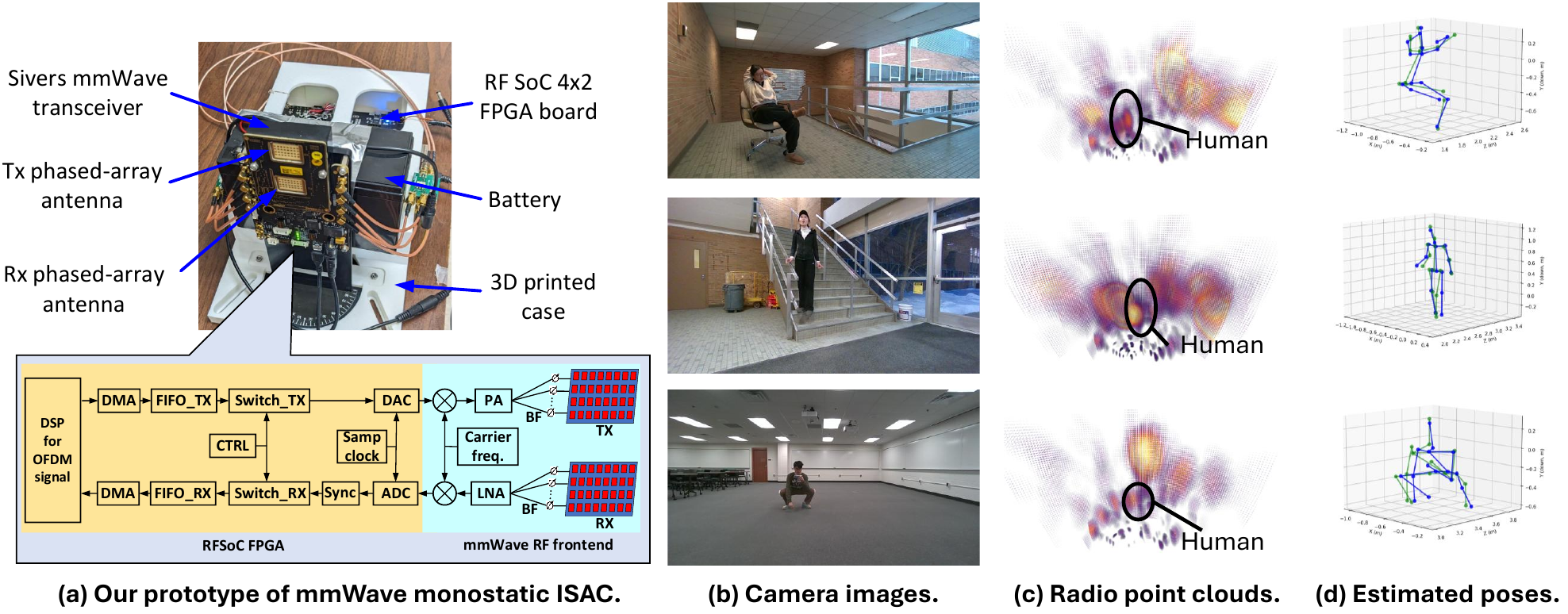}
\caption{An example of an mmWave monostatic ISAC system, the generated radio point clouds, and its downstream application for pose estimation. In (d), the green skeleton denotes the depth-camera ground truth, while the blue skeleton shows the ISAC-estimated pose.}
\label{fig:mmwave_isac2}
\end{figure*}

\item 
\textbf{Doppler:}  In a monostatic configuration, coherent detection enables the estimation of Doppler frequency shifts caused by object motion. In OFDM-based ISAC systems, when OFDM symbols are transmitted at regular time intervals and a sufficient number of symbols are available, Doppler frequencies can be estimated through an FFT operation.
When the number of available OFDM symbols is limited, the velocity of an object can instead be estimated from the phase variation of CIR taps between consecutive OFDM symbols:
$
v \approx \frac{\lambda}{4\pi \tau}
\angle \left(\frac{h_n(t+1)}{h_n(t)}\right)
$,
where \(h_n(t)\) denotes the \(n\)-th CIR tap at time slot \(t\), \(\lambda\) is the signal wavelength, and \(\tau\) is the time interval between two OFDM symbols.

\item
\textbf{Angular Estimation:}  
Angular estimation is primarily determined by the antenna array geometry rather than the transmitted waveform. 
Therefore, the angular estimation techniques used in ISAC systems are similar to those used in conventional radar systems such as FMCW radars.
If antenna elements are arranged along a single axis, the system can estimate the azimuth angle of incoming reflections. 
If antennas are arranged over a two-dimensional plane as shown in Fig.~\ref{fig:cir2}, both azimuth and elevation angles can be estimated. 
By combining range, Doppler, azimuth, and elevation measurements, the system can construct 4D sensing data representing the spatial and motion characteristics of objects in the environment.

\end{itemize}

\section{Monostatic ISAC on mmWave Device}

In this section, we study the monostatic ISAC on individual mmWave communication devices.
MmWave communication devices typically employ phased-array antennas for both transmission and reception. The directional beamforming capability of phased arrays, together with the short wavelength at mmWave frequencies, naturally provides significant isolation between the transmitter and receiver. This isolation makes it feasible to operate in full-duplex mode for monostatic sensing.

\begin{table}[t]
\centering
\caption{The parameters of our monostatic ISAC device.}
\begin{tabular}{|ll|}
\hline
\textbf{Hardware parameters} & \\
\hline
Sampling rate & 1.2288~GSPS \\
Number of Tx antennas & 16 \\
Number of Rx antennas & 16 \\
Center frequency & 60~GHz \\
Transmission power & 20~dBm \\
\hline
\textbf{Communication parameters} & \\
\hline
Waveform & OFDM \\
FFT points & 1024 \\
Number of valid subcarriers & 900 \\
Cyclic prefix length & 276 \\
OFDM symbol duration & 1.057~$\mu$s \\
Number of OFDM symbols per frame & $16\times16 = 256$ \\
Supporting protocols & 5G and Wi-Fi \\
\hline
\textbf{Sensing parameters} & \\
\hline
Detection time of a frame & 67~$\mu$s \\
Number of frames per second & 10 \\
Practical detection range & 10~m \\
Number of effective antennas (horizontal) & 8 \\
Number of effective antennas (elevation) & 4 \\
Horizontal antenna spacing & 0.5~wavelength \\
Elevation antenna spacing & 0.5~wavelength \\
Horizontal field of view & [-60$^\circ$, 60$^\circ$]\\
Elevation field of view & [-30$^\circ$, 30$^\circ$]\\
\hline
\end{tabular}
\label{tab:system_params}
\end{table}

\subsection{Experimental Validation}
To validate this approach, we built a prototype for mmWave OFDM communication and monostatic sensing as shown in Fig.~\ref{fig:mmwave_isac2}(a). 
Table~\ref{tab:system_params} lists the system parameters.
The device hardware consists of two primary commercial off-the-shelf (COTS) modules:
(i) an AMD/Xilinx RFSoC 4x2 FPGA board, and 
(ii) a Sivers mmWave transceiver equipped with Tx/Rx phased-array antennas.
The RFSoC FPGA implements the OFDM signal processing pipeline that supports both 5G and Wi-Fi physical-layer protocols, while the mmWave transceiver handles 60\,GHz radio transmission and reception. 
Within the FPGA, the data transmission and reception pipelines are jointly optimized and carefully calibrated to ensure timing and clock alignment for monostatic sensing. 
The phased-array antenna control is integrated into the processing pipeline to synchronize beam steering with signal transmission.


To measure the strength of self-interference, we placed a triangular reflector in front of the mmWave device at a distance of 1\,m, with both of its transmitter and receiver beams directed toward the reflector. We then collected CIR measurements, in which the self-interference appears at the first channel tap. The results show that the self-interference strength is comparable to the signal reflected from the reflector, indicating that the self-interference remains within an acceptable range.


\begin{figure*}
    \centering
    \includegraphics[width=\linewidth]{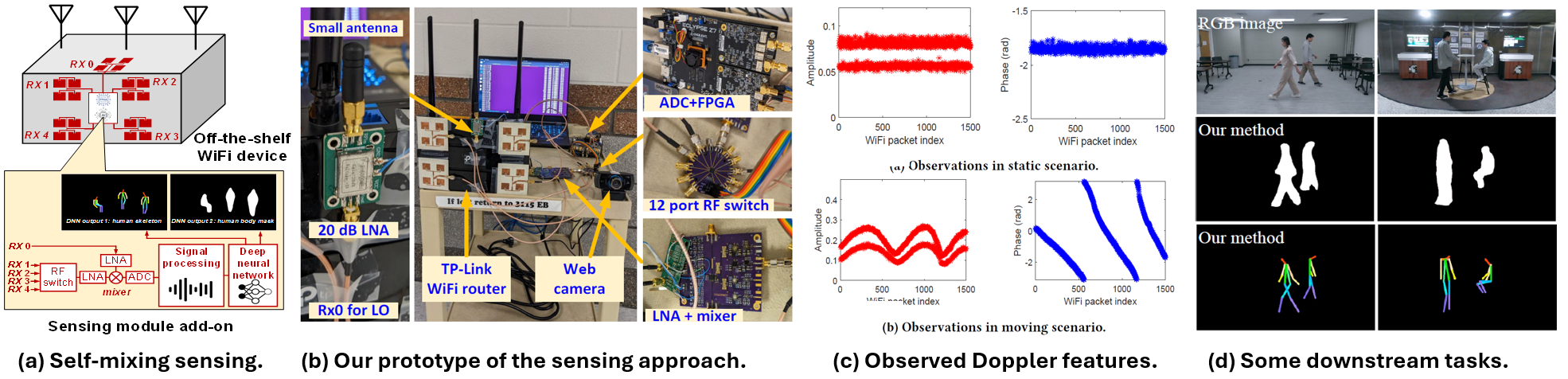}
    \caption{An example of the self-mixing approach that uses the OFDM communication signal for sensing applications.}
    \label{fig:selfmixing}
\end{figure*}

\subsection{RF Sensing Data}

The mmWave device is equipped with two phased-array antennas: one for transmitter and one for receiver.
Using the two phased-array antennas, the prototype can collect spatially resolved CIR measurements across multiple beam directions. 
For each transmitted OFDM frame, the receiver first estimates the channel coefficients of all subcarriers and then converts the frequency-domain channel measurements into time-domain delay profiles.
By combining the delay profiles obtained under different beam directions, the device can estimate both range and angular information of reflected signals.  
Successive frames collected over time enable Doppler estimation through phase evolution across CIR measurements. 
As a result, the device can generate 4D sensing data consisting of range, velocity, azimuth, and elevation information, which can be further processed to produce point-cloud representations of the surrounding environment.

\subsection{Downstream Applications}

Monostatic ISAC can support a wide range of applications, from human activity recognition to 3D scene reconstruction \cite{Song2026Rascene}. Fig.~\ref{fig:mmwave_isac2}(b-d) illustrates its application to human pose estimation. Specifically, subfigure (b) shows the camera images as references, (c) presents the radio point cloud generated by the device, and (d) shows the estimated poses produced by a well-trained DNN model supervised by a depth camera.
In Fig.~\ref{fig:mmwave_isac2}(b), the green skeleton denotes the ground-truth pose obtained from the depth camera, while the blue skeleton represents the pose estimated from the ISAC measurements.

\section{Monostatic ISAC on Sub-6GHz Device}

In this section, we study monostatic ISAC on sub-6,GHz communication devices. Unlike mmWave devices, sub-6\,GHz systems typically employ omnidirectional antennas. As a result, self-interference can dominate the received signal and potentially saturate the ADC, preventing the receiver from capturing weak reflections from surrounding objects. 
In what follows, we discuss three approaches for managing self-interference.

\subsection{Self-Interference Cancellation}

Self-interference cancellation techniques developed for full-duplex communications provide a promising solution to this monostatic ISAC problem. By combining analog and digital cancellation methods, modern systems can achieve more than 70\,dB of self-interference suppression \cite{mohammadi2023comprehensive}.
From a system perspective, although such cancellation levels have been demonstrated in research prototypes, the large-scale deployment of full-duplex communication has progressed slowly in practical systems and in 3GPP standardization due to additional engineering challenges beyond theoretical cancellation limits \cite{mohammadi2023comprehensive}. This raises the question of whether self-interference cancellation techniques can enable monostatic ISAC in real systems.

We believe the answer is affirmative. This is because there is a fundamental difference between \textit{full-duplex communication} and \textit{full-duplex sensing}. 
In monostatic ISAC, full-duplex is for sensing, not for communication. 
The requirements for full-duplex sensing are generally much less stringent than those for full-duplex communication. In full-duplex communication, the receiver must decode extremely weak signals from a remote transmitter in the presence of strong self-interference. In contrast, full-duplex sensing primarily aims to estimate CIR of reflected signals, a task that is inherently more tolerant to residual interference and noise.

\subsection{Self-Mixing RF Down-Conversion}
Another approach to enabling monostatic ISAC is a self-mixing RF down-conversion RF architecture. 
This approach was inspired by the architecture of FMCW radar. 
In FMCW radar systems, the received radio signals are directly mixed with a copy of the transmitted FMCW signal in the RF domain, generating an intermediate-frequency (IF) signal whose spectral components reveal object distances.
A similar architecture can be applied to communication devices for sensing, as illustrated in Fig.~\ref{fig:selfmixing}(a). 
Compared to FMCW radar, the main difference is that the transmitted signal is an OFDM waveform rather than an FMCW waveform. 

By self-mixing the transmitted OFDM signal with the received signal in the RF domain, an intermediate-frequency (IF) signal can also be generated and used to extract environmental information. Unlike the self-mixing of FMCW signals, which directly reveals the ranging information of surrounding objects, self-mixing OFDM signals does not inherently provide object distance information. Nevertheless, the resulting IF signal still contains Doppler information associated with moving objects in the environment. Through appropriate signal processing of the IF signal, this Doppler information can be extracted to characterize object motion \cite{song2024siwis}. Although ranging information is unavailable, Doppler information (together with angular information) can still support many downstream applications, such as gesture recognition and human activity recognition.

\noindent
\textbf{Implementation.}
Fig.~\ref{fig:selfmixing}(b) shows our prototype of an OFDM self-mixing system.
The prototype is based on a COTS TP-Link Wi-Fi router operating on channel~44. 
It has no modification on the Wi-Fi router, but adds a sensing hardware for monostatic sensing.
The add-on sensing hardware comprises a small dipole antenna (RX0), four patch antennas (RX1--RX4), an RF switch, a custom-designed RF PCB, an ADC daughterboard, and an ECLYPSE Z7 FPGA board. 
The RF PCB was designed using Analog Devices' HMC951A mixer and HMC717A low-noise amplifier (LNA), and fabricated on an OSH Park FR408 substrate. 
The four patch antennas were designed using HFSS electromagnetic simulation software and fabricated on Rogers RO4350B substrate.
The FPGA controls the RF switch and the ADC daughterboard to sample the received signals at 10~MSps. 
The sampled data are transmitted to a laptop via Ethernet for offline signal processing. 


\noindent
\textbf{Experimental Validation.}
Fig.~\ref{fig:selfmixing}(c) shows the measured amplitude and phase of the sensing signal on our prototype under two scenarios: (i) all objects are static, and (ii) a single object moves toward the prototype. The results indicate that in a static environment, the sensing signal remains unchanged, whereas the amplitude and phase vary according to the speed and direction of the moving object. This behavior confirms that the sensing signal reliably encodes Doppler information.

\noindent
\textbf{Downstream Applications.}
Fig.~\ref{fig:selfmixing}(d) illustrates two downstream applications of the monostatic ISAC system: human mask segmentation and skeleton estimation. A DNN model was designed and trained using data from the camera co-located with the device, and evaluated in previously unseen scenarios. The results demonstrate that, even without ranging information, this monostatic ISAC approach can detect moving objects with fine-grained detail.

\subsection{Precise Tx/Rx RF Switching Control}

Unlike the previous three approaches that enable simultaneous transmission and reception, this method relies on rapid alternation between signal transmission and reception, mimicking the operation of a pulse radar. In this scheme, the device does not transmit and receive at the same time. Instead, once the device finishes transmitting an OFDM signal for a data packet, it quickly switches to receive mode to capture the tail of the transmitted signal. The switching interval is tightly controlled to allow inference of the signal propagation delay profile.

Thanks to advances in semiconductor technology, modern RF switches can operate at nanosecond-level speeds with deterministic delay. As a result, the receiver can capture both the tail of the transmitted packet and its multipath reflections. By analyzing the received signal together with the known transmitted waveform, the system can extract multipath information and estimate both range and Doppler parameters. This approach enables communication devices to achieve pulse radar-like sensing functionality without requiring continuous full-duplex operation.

In practice, the main advantage of this approach is its hardware simplicity. However, it is only suitable for wideband signal systems and cannot detect nearby objects, being effective only for targets beyond a certain minimum distance.

\section{OFDM for Monostatic ISAC}

In this section, we discuss the suitability of OFDM waveform for sensing and present methods to improve its sensing range.

\subsection{OFDM versus FMCW}

In general, \textit{environment-independent} waveforms are preferred for communication, while \textit{environment-dependent} waveforms are better suited for sensing. Accordingly, OFDM waveforms are widely regarded as ideal for communication, whereas FMCW waveforms are considered well-suited for sensing \cite{kumar2026amalgamated}. While this distinction generally holds, we argue that OFDM waveforms can also support all sensing functionalities. As demonstrated earlier, both OFDM and FMCW waveforms are capable of providing ranging, Doppler, and angular information of objects.

Nevertheless, the two waveforms differ in terms of energy efficiency, hardware complexity, and computational cost.
\begin{itemize}[leftmargin=0.15in]
\item 
\textbf{Energy Efficiency:} OFDM signals have a high peak-to-average power ratio (PAPR), requiring power amplifiers with a larger dynamic range, whereas FMCW signals have constant amplitude. As a result, FMCW devices are generally more power-efficient than OFDM devices.

\item 
\textbf{Hardware Requirements:} OFDM-based sensing needs high-speed ADCs to sample the full signal bandwidth for CIR estimation, while FMCW systems typically down-convert the RF signal to an intermediate frequency and can use low-rate ADCs. Additionally, OFDM receivers require two ADC channels for I/Q sampling, whereas FMCW systems can operate with a single ADC.

\item 
\textbf{Computation Requirements:} OFDM systems must process wideband signals (e.g., 100\,MHz), leading to higher digital signal processing complexity compared with FMCW systems that handle narrower bandwidths (e.g., 10\,MHz).
\end{itemize}

\subsection{Sensing Range}

Although monostatic ISAC enables ranging and Doppler estimation, one challenge is the achievable sensing range. Compared with bistatic sensing, where signals travel only one-way from transmitter to receiver, monostatic sensing requires the signal to travel a two-way propagation path (transmit-reflect-receive). In addition, the high PAPR of OFDM signals further limits the achievable sensing range.

One approach to improving sensing reliability and extending sensing range is to exploit the temporal diversity gain of multiple CIR measurements. This idea relies on two observations. First, since the transmitter and receiver are co-located on the same device, the receiver knows the payload data of every OFDM symbol and can therefore estimate the CIR from every OFDM symbol in a data packet. Second, with increasing bandwidth in future 6G and next-generation Wi-Fi systems, the duration of an OFDM symbol becomes very short, typically on the order of microseconds.

For example, in 5G NR with 100\,MHz bandwidth and 30\,kHz subcarrier spacing, the duration of one OFDM symbol is approximately 34\,$\mu$s. During this interval, an object moving at a speed of 10\,m/s travels less than 0.344\,mm, which is much smaller than the radio wavelength. Therefore, the sensing signal can be considered approximately constant across multiple OFDM symbols.

Therefore, the sensing reliability and range can be enhanced by jointly combining a group of consecutive OFDM symbols in a data packet.
Specifically, the sensing measurements can be modeled as
\(
x_i = s + w_i
\),
where $x_i$ denotes the $i$th observation, $s$ is the underlying sensing signal, and $w_i$ represents independent noise. By combining multiple OFDM observations through
\(
\sum_{i=1}^{N} x_i
\),
the signal-to-noise ratio (SNR) can be improved. 
Aggregating $N$ measurements increases the SNR by $10\log_{10}(N)$ dB compared with a single measurement, thereby improving sensing reliability and extending the effective sensing range.

\subsection{OFDM-MIMO for Monostatic ISAC}

Multiple-input multiple-output (MIMO) antenna arrays play a critical role in both monostatic and bistatic ISAC systems. MIMO provides spatial degrees of freedom that enable the estimation of angular information in both azimuth and elevation directions.
The theoretical spatial resolution of an antenna array is determined by the array aperture. However, the achievable angular estimation accuracy also depends on factors such as the number of antennas, their geometric arrangement, and the antenna structure. 
By integrating antenna arrays into communication devices, OFDM-based monostatic ISAC systems can perform four-dimensional sensing that includes range, Doppler, azimuth, and elevation measurements.

\section{Monostatic ISAC Applications}

Monostatic ISAC has the potential to provide radar-like sensing capabilities using existing wireless communication infrastructure. By leveraging OFDM communication signals that are already transmitted in cellular and Wi-Fi networks, monostatic ISAC can perform environmental sensing through hardware, spectrum, and/or waveform sharings. This capability makes monostatic ISAC a promising technology for enabling large-scale sensing applications in future 6G wireless networks.

\subsection{Monostatic ISAC on Stationary Devices}

Cellular base stations and Wi-Fi routers are typically deployed at fixed locations within wireless network infrastructure. Their stationary positions and continuous operation make them ideal platforms for environmental sensing using monostatic ISAC. Individual communication devices can potentially perform many sensing tasks that are traditionally associated with camera-based computer vision systems in indoor and urban environments. These tasks include object detection, semantic and instance segmentation, pose estimation, object tracking, action recognition, depth estimation, scene understanding, and 3D reconstruction. Below, we discuss several representative applications of monostatic ISAC.

\textbf{Human Activity Recognition.}  
Monostatic ISAC devices can capture RF reflections from the human body and track motion dynamics over time. These signals enable the inference of fine-grained human activities and body movements without requiring wearable sensors or cameras. As a result, monostatic ISAC can support a wide range of human-centric sensing applications, including 3D human pose estimation, activity recognition, gesture recognition, fall detection for elderly care, human-computer interaction, and human re-identification.

\textbf{Smart Building Sensing.}  
Stationary communication devices can also monitor indoor environments and building occupancy using RF reflections from surrounding objects and people. Unlike camera-based solutions, RF sensing can operate without visible-light illumination and can preserve user privacy because it does not capture visual imagery. Consequently, monostatic ISAC enables privacy-friendly smart building applications such as room occupancy detection, people counting, intrusion detection, and intelligent energy management for heating, ventilation, and lighting systems.

\textbf{Occlusion-Resilient Sensing.}  
Monostatic ISAC enables through-occlusion sensing by leveraging radio signals that can penetrate or diffract around certain materials such as walls, furniture, smoke, or foliage. When a communication device transmits OFDM RF signals, reflections from objects that are not directly visible can still return to the co-located receiver through multipath propagation. By analyzing the resulting CIR, Doppler shifts, and angular information, the device can infer the presence, motion, and spatial structure of occluded objects. This capability allows monostatic ISAC systems to perform sensing tasks even when line-of-sight is blocked, making them valuable for applications such as indoor monitoring, search-and-rescue operations, security surveillance, and autonomous navigation in cluttered environments.

\subsection{Monostatic ISAC on Mobile Devices}

Mobile communication devices such as vehicles, drones, and smartphones can also benefit from monostatic ISAC capabilities. 
Because these devices move through different environments, they enable dynamic sensing and mobile perception while simultaneously supporting wireless communication. For example, vehicles equipped with monostatic ISAC can perform environmental sensing to support applications such as vehicle-to-vehicle sensing, pedestrian detection, collision avoidance, cooperative perception, and traffic monitoring, thereby improving autonomous driving safety. Similarly, unmanned aerial vehicles can leverage monostatic ISAC for obstacle detection, autonomous navigation, drone-to-drone coordination, search-and-rescue operations, and airspace monitoring. 

Compared to stationary devices, mobile devices can leverage its spatial trajectory to enrich the information of detection by combining the sensing data collected at different time moments or at different locations along their trajectory. 
The multi-frame fusion can significantly improve the sensing performance for some applications such as SLAM and 3D scene reconstruction \cite{Song2026Rascene}. 
Below, we present two approaches for multi-frame data fusion for mobile devices.

\begin{itemize}[leftmargin=0.15in]
\item 
\textbf{Phase-Level Multi-Frame Fusion.}  
For mobile sensing devices, multi-frame measurements can be combined to improve sensing resolution. 
If the device location can be obtained with sufficiently high accuracy, i.e., on the order of a small fraction of the signal wavelength, synthetic aperture radar (SAR) techniques can be applied. 
In this case, measurements collected from different device positions can be coherently fused in the phase domain to synthesize a larger virtual antenna aperture, enabling high-resolution SAR imaging \cite{moro2024exploring}.

\item 
\textbf{Feature-Level Multi-Frame Fusion.}  
In many cases, achieving wavelength-level localization accuracy for mobile devices is not possible. 
Then, alternative multi-frame fusion techniques can still exploit multiple observations to improve sensing performance.
Instead of phase-level fusion, feature-level multi-frame fusion methods can be applied. 
For example,~\cite{Song2026Rascene} presents a multi-frame fusion framework in which each RF sensing frame is first transformed into a latent representation using a neural encoder. 
Features from multiple frames are then aggregated in the latent space before being decoded back into the spatial domain to generate improved sensing outputs. 
Although the performance gain is generally smaller than that achieved by SAR, carefully designed multi-frame fusion methods can still provide significant improvements in detection and reconstruction accuracy.

\end{itemize}

\section{Research Directions}

In this section, we discuss several research directions for advancing monostatic ISAC and promoting its deployment in future 6G networks.

\subsection{Hardware Support for Monostatic ISAC}

Future wireless systems may incorporate dedicated hardware features to support ISAC functionalities, such as improved antenna architectures, integrated sensing pipelines, and efficient full-duplex transceiver designs. In particular, the following research directions warrant further investigation.

\begin{itemize}[leftmargin=0.15in]

\item 
\textbf{Full-Duplex Operation:}  
Enabling efficient full-duplex operation is one of the key challenges for monostatic ISAC systems. In both FR2 (mmWave) and FR1 (sub-6\,GHz) bands, strong transmitter leakage can overwhelm the receiver and severely degrade sensing performance. Achieving sufficient self-interference suppression requires careful co-design of antenna isolation, analog cancellation, and digital cancellation techniques \cite{yang2024parallel,xiao2025ris}.

\item 
\textbf{Advanced Antenna and Beamforming:}  
Angular resolution is a critical factor for RF sensing. It is determined by the antenna aperture, the number of antenna elements, and the array geometry. In addition, antenna characteristics such as radiation efficiency and radiation patterns directly affect the achievable sensing range. Therefore, an important research direction is the design of advanced antenna architectures and beamforming strategies that are jointly optimized for both communication and sensing.

\item 
\textbf{RF Calibration:}  
RF front-end components often exhibit imperfections in phase and gain control, particularly in mmWave circuits. Accurate angular estimation and high-resolution sensing rely on precise phase measurements across antenna arrays. Consequently, efficient RF calibration techniques are required to compensate for hardware impairments. Ideally, calibration procedures should incur low overhead and be performed infrequently while maintaining long-term stability.

\end{itemize}

\subsection{AI for Monostatic ISAC}

AI will play a key role in RF sensing, including monostatic, bistatic, and other ISAC architectures \cite{aldirmaz2025comprehensive}. Because RF signals interact with the environment through complex effects such as multipath propagation, scattering, and reflections, accurate analytical modeling is challenging. Deep learning can instead learn nonlinear mappings directly from RF measurements to sensing outputs, enabling end-to-end optimization of the sensing pipeline. Despite recent progress, several research challenges remain.

\textbf{Robust Feature Extraction:}
RF measurements are often corrupted by noise, interference, multipath propagation, and hardware imperfections, which degrade sensing performance. Effective feature extraction methods must therefore be robust to such impairments while remaining computationally efficient for real-time deployment. In addition, RF propagation varies significantly across environments, making environment-invariant representations essential for reliable generalization.

\textbf{Deep Learning Architectures:}
CNN-based models are widely used to process RF representations such as CSI, spectrograms, and range–Doppler maps, while emerging architectures including diffusion and flow-based models show promise for RF signal modeling. However, models trained in one environment often fail to generalize to new scenarios. Incorporating wireless domain knowledge into model design and developing improved training strategies, including better loss functions and multi-task learning, remain important research directions.

\textbf{Unsupervised Learning:}
Most RF sensing systems rely on supervised learning with labeled datasets obtained from auxiliary sensors such as cameras or LiDAR. However, large-scale labeled datasets are expensive to collect. Unsupervised and self-supervised learning offer promising alternatives by learning meaningful representations directly from unlabeled RF data, reducing the dependence on costly annotations.

\textbf{AI for Multi-Modal Sensing:}
RF sensing is robust to occlusion, lighting changes, and adverse weather but typically provides lower spatial resolution than optical sensors. Integrating RF sensing with complementary modalities such as cameras or LiDAR through AI-driven multi-modal fusion can improve perception accuracy and robustness.

\textbf{Large RF Models for 6G:}
Inspired by foundation models in NLP and computer vision, large-scale RF models trained on massive RF datasets may support multiple sensing tasks such as activity recognition, localization, and environmental mapping. Combined with monostatic ISAC-enabled devices, such models could enable large-scale distributed sensing platforms in future 6G networks.

\section{Concluding Remarks}

Monostatic ISAC represents a promising paradigm for enabling environmental perception using wireless communication infrastructure. 
By leveraging OFDM communication signals together with emerging full-duplex technologies, monostatic ISAC can provide radar-like sensing capabilities without requiring dedicated sensing hardware or additional spectrum resources. 
With continued advances in RF hardware, signal processing techniques, and AI models, monostatic ISAC is expected to play an important role in future 6G wireless systems.


\bibliographystyle{IEEEtran}
\bibliography{references}
\end{document}